%% file: v10.tex
\documentclass[aps,prl,superscriptaddress,twocolumn,floatfix]{revtex4-2}
\usepackage[abs]{overpic}
\usepackage{graphicx,graphics}
\usepackage{dcolumn}
\usepackage{amsmath,amssymb,amsfonts}
\usepackage{wasysym}
\usepackage{latexsym,verbatim}
\usepackage{bm}
\usepackage[dvipsnames]{xcolor}
\usepackage[framemethod=default]{mdframed}
\usepackage[breaklinks=true,colorlinks,citecolor=blue,linkcolor=blue,urlcolor=blue]{hyperref}
\usepackage[makeroom]{cancel}
\usepackage{fancybox}
\usepackage{ulem}
\usepackage{tikz}

\newcommand{\be}{\begin{eqnarray}}
\newcommand{\ee}{\end{eqnarray}}
\newcommand{\nn}{\nonumber\\}

\newcommand{\sgn}{\text{sgn}}

\begin{document}

\title{Moir\'{e} driven edge reconstruction in Fractional quantum anomalous Hall states}

\author{Feng Liu}
\author{Hoi Chun Po}
\author{Xue-Yang Song}
\affiliation{Department of Physics, The Hong Kong University of Science and Technology, Clear Water Bay, Hong Kong, China}
\affiliation{Center for Theoretical Condensed Matter Physics, The Hong Kong University of Science and Technology, Clear Water Bay, Hong Kong, China}

\begin{abstract}
We investigate fractional edge modes in moir\'{e} fractional quantum anomalous Hall states, focusing on the role of lattice momentum conservation and umklapp scattering.
For the hierarchical $\nu=2/3$ state, we show that, for a class of microscopic edge realizations, moir\'{e}-enabled umklapp processes can stabilize the Kane-Fisher-Polchinski fixed point even in the absence of disorder.
Our results illustrate how lattice momentum constraints can qualitatively reshape the interaction structure and low-energy behavior of fractional edge modes. 
The study of Umklapp processes in edge reconstruction serves as a crucial bridge to understanding thermal and electrical transport in the hierarchical fractional quantum anomalous Hall states found in lattice systems of quantum simulators.
\end{abstract}

\maketitle

{\it Introduction}---Moir\'{e} systems have emerged as a versatile platform for exploring strong electronic correlations in two dimensions.
Owing to the large moir\'{e} unit cell and the narrow electronic bands, a wide variety of correlated phases have been observed or proposed in twisted van-der-Waals hetero structures, including correlated insulators~\cite{cao2018correlated}, superconductors~\cite{cao2018unconventional,supercondcutivtiy2025dean,xia2026,tingxin2025}, and (fractional) quantum anomalous Hall (FQAH) states~\cite{nuckolls2020strongly,cai2023signatures,zeng2023thermodynamic,park2023observation,lu2023fractional}, following directions suggested by theoretical works~\cite{zhang2019nearly,ledwith2020fractional,repellin2020chern,abouelkomsan2020particle,wilhelm2021interplay,wu2019topological,yu2020giant,devakul2021magic,li2021spontaneous,crepel2023fci,zhou2024fractional,zhang2024moore,dong2024anomalous,dong2024theory,kim2025topological}.
Among the remarkable zoo of correlated phases, FQAH are particularly intriguing with intrinsic topological orders, anyon excitations and associated phase transitions~\cite{song_2024phase,song_2024_intertwined,song_2024density,shi2025doping}. FQAH hosts robust edge state transport, described by chiral Luttinger liquids in an ideal setting~\cite{wen_ll}. Multiple factors such as disorder and confining potential are shown to strongly renormalize the edge states, leading to edge reconstruction and distinct transport signatures~\cite{kane1995impurity,wang2013edge,tam2026quantized,manna2025multiple,manna2024shot}, which constitutes a central topic under extensive investigation in condensed matter physics. 
In the FQAH phase on moir\'{e} superlattices, factors unique to moir\'{e}-scale physics may affect the edge reconstruction, which is the main focus of this Letter. 

The theoretical description of strongly correlated moir\'{e} materials is complicated by the presence of extremely large moir\'{e} supercells, which introduce a vast number of electronic degrees of freedom. 
This complexity makes fully microscopic many-body approaches computationally prohibitive, motivating the search for simplified low-energy effective theories that retain the essential interaction driven physics.
One fruitful approach is provided by network and coupled-wire descriptions, in which a two dimensional system is viewed as an array of interacting one dimensional channels. 
Experimental observations of highly anisotropic transport in moir\'{e} materials further indicates the presence of arrays of one-dimensional channels and quantum-network structures ~\cite{wang2022one}.
Within this framework, electron-electron interactions are naturally incorporated through Luttinger liquid theory, and correlated phases emerge from interwire scattering processes~\cite{kane2002fractional,teo2011luttinger}.
In this context, coupled-wire and network models have provided a powerful lattice realization of quantum Hall physics~\cite{wu2019coupled,wang2022one,fujimoto2022perfect}, successfully capturing integer Chern insulators and their associated edge modes in moir\'{e} systems~\cite{hsu2023general}.

Whether this correspondence can be generalized to FQAH/fractional Chern insulators (FCI) remains open.
In particular, embedding fractional edge theories into lattice realizations modifies the momentum constraints inherent to the coupled-wire construction, allowing umklapp processes and inter-edge tunneling channels that are forbidden when the momentum scale is set by a real magnetic field.
From this perspective, a central question is how such lattice-enabled umklapp processes affect the structure and stability of interacting fractional edge modes.

In this work, we investigate these effects in detail for the $\nu=2/3$ FQAH states, motivated by the experimental signatures of FQAH states in semiconductor and graphene-based moir\'{e} platforms~\cite{zeng2023thermodynamic,lu2023fractional}.
Using a coupled-wire construction adapted to moir\'{e} systems, we show that the hierarchical $\nu=2/3$ state admits multiple microscopic realizations that share the same bulk topological order but differ in the momentum structure of their edge electron operators. 
For a class of realizations natural to lattice systems, lattice-enabled umklapp scattering introduces a relevant but otherwise forbidden inter-edge tunneling channel, and stabilizes the Kane-Fisher-Polchinski (KFP) fixed point even in the absence of disorder.
These results identify a lattice-enabled mechanism---concretely illustrated within a coupled-wire framework---by which edge modes in lattice systems acquire interaction channels that are absent in the continuum.
In this light, our results offer an alternative interpretation of recent local probe measurements of fractional Chern insulators~\cite{ji2024local}, suggesting that lattice-enabled umklapp processes can qualitatively affect the equilibration and stability of fractional edge modes without invoking strong disorder.

\begin{figure}[t!]
\begin{overpic}[width=0.8\linewidth]
            {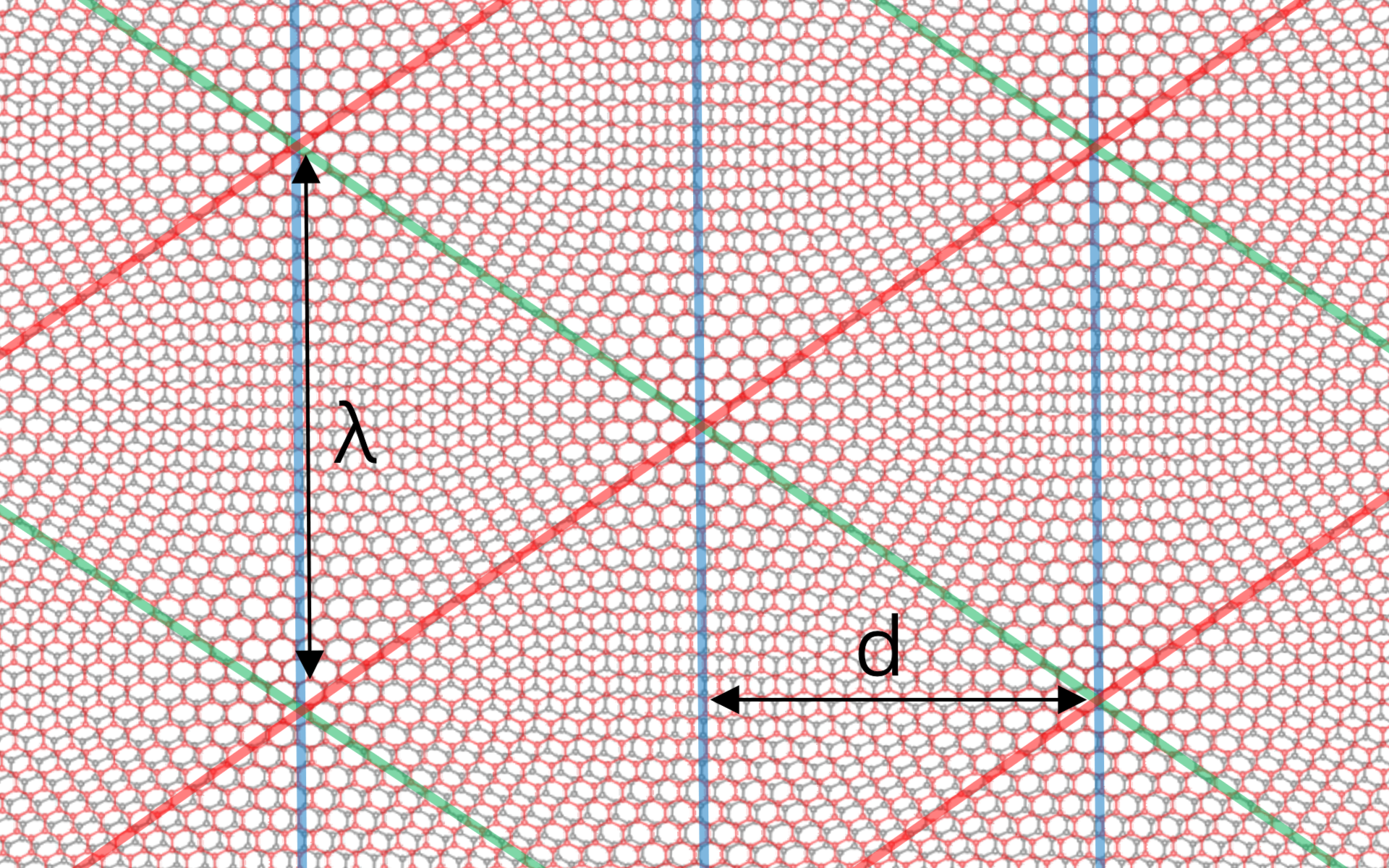}
        \end{overpic}
    \caption{Schematic illustration of a quantum-wire network defined on a moir\'{e} superlattice. 
    The underlying moir\'{e} pattern has wavelength $\lambda = a_0/[2\sin(\theta/2)]$, set by the twist angle $\theta$.
    Three sets of parallel quantum wires are highlighted in distinct colors, forming a triangular network; within each set, neighboring wires are separated by a distance $d=\sqrt{3}\lambda/2$.
    } 
    \label{fig:1}
\end{figure}

{\it Coupled-wire construction}---We briefly review the coupled-wire construction for quantum Hall states~\cite{teo2011luttinger,kane2002fractional}.
We here adopt a minimal single-channel Luttinger-liquid framework and assume spin and valley polarization, which captures the essential interaction effects~\cite{hsu2023general,chou2019superconductor}.
Extensions to include extra internal degrees of freedom can be formulated within the same framework but not required for the present analysis.
%

In this approach, a two dimensional system is viewed as an array of parallel one dimensional quantum wires indexed by $j$, each described at low energies by a Luttinger liquid.
The electron operators can be bosonized as
\be
\psi_{R/L,j}(x)
&=&\frac{\kappa_j}{\sqrt{2\pi a}}\exp(i(k_{Fj}^{R/L}x+\varphi_j(x)\pm\theta_j(x))),
\ee
where $a$ is a short-distance cutoff, $\kappa_j$ is a Klein factor. 
and $k_{Fj}^{R/L}=bj\pm k_F^0$, with $b$ encoding the momentum offset induced by the background magnetic field in the quantum Hall setting~\cite{kane2002fractional}.
%

Interwire interactions are constructed from local products of single electron operators~\footnote{While intra-set scattering processes play a central role in the coupled-wire construction discussed in the main text, scattering processes between different wire sets are subject to more stringent RG relevance conditions and therefore less relevant at low energies~\cite{hsu2023general}; accordingly, we focus on intra-set processes in the present work.}, with the most general form~\cite{teo2011luttinger}
%
%
%
\be
\mathcal{O}_j^{\{s_p^L,s_p^R\}}(x)&=&\prod_p (\psi^R_{j+p}(x))^{s_p^R}(\psi^L_{j+p}(x))^{s_p^L},
\ee
%
where $j$ labels the wire, $p$ its relative displacement, and the integers $s_p^{R/L}$ specify the numbers of right-/left-moving electron operators (negative values denote Hermitian conjugation).
%
%
%
Charge conservation requires $\sum_{p}(s_{p}^R+s_p^L)=0$, while momentum conservation imposes the condition~\cite{kane2002fractional,teo2011luttinger}
\be
k_F^0\sum_p(s_p^R-s_p^L)+b\sum_p p(s^R_p+s^L_p)=0.
\ee

To adapt the coupled-wire construction to moir\'{e} systems to describe interaction driven Chern insulating phases, we interpret the parameter $b$ as an effective momentum bookkeeping scale that organizes the relative Fermi momenta of neighboring wires. 
Rather than assuming a physical magnetic field or a pre-existing Chern band, we use this bookkeeping scheme in reverse: we ask what momentum structure is required if the system realizes a quantum-Hall-like state, and whether such a structure can be supplied by the moir\'{e} lattice.
%

In continuum quantum Hall systems, an analogous role is played by the magnetic field, which fixes momentum offsets between wires and constrains multi-wire tunneling processes~\cite{kane2002fractional,teo2011luttinger}.
In Chern insulators~\cite{ledwith2020fractional,parameswaran2012fractional}, a similar dictionary is often expressed through $l_B^2\rightarrow\langle \Omega \rangle$, where $l_B$ is the magnetic length and $\langle\Omega\rangle$ is the average Berry curvature of the Chern band.
%
We invoke this correspondence as a geometric dictionary for converting the moir\'{e} lattice scale into the momentum offsets entering the coupled-wire description.

Within this framework, the quantity $2\pi l_B^2$ is identified with the area of an effective unit cell defined by commuting translation operators.
For moir\'{e} materials, this effective unit cell coincides with the moir\'{e} superlattice cell, yielding~\cite{ledwith2020fractional}
\be
2\pi l^2_B =\frac{\sqrt{3}}{2}\lambda^2,
\ee
where $\lambda=a_0/(2\sin(\theta/2))$ is the moir\'{e} lattice constant determined by the twist angle $\theta$.
In the quantum-wire network representation of the moir\'{e} pattern shown in Fig.~\ref{fig:1}, neighboring wires are separated by a distance $d=\sqrt{3}\lambda/2$, as discussed in Ref.~\cite{hsu2023general}.
This separation translates into a momentum offset between neighboring wires, $\Delta k=d/l_B^2=2\pi/\lambda$, in direct analogy with the continuum quantum Hall systems~\cite{asasi2021equilibration}.
Comparing with the bosonized form of the fermion operators, where the Fermi momenta are parameterized as $k_{Fj}^{R/L}=bj\pm k_F^0$, we therefore identify 
\be
b=2\pi/\lambda
\ee
%
The momentum scale $b$ required by the coupled-wire description is not imposed externally, rather naturally supplied by the moir\'{e} superlattice, coinciding with a reciprocal lattice vector $2\pi/\lambda$.
This is consistent with earlier coupled-wire constructions with $b=0$~\cite{hsu2023general}, while making explicit the role of lattice-enabled umklapp processes.
%

%
In the continuum fractional quantum Hall setting, tunneling processes between edge modes at fillings such as $\nu=2/3$ are typically rendered irrelevant by momentum mismatches, which generate rapidly oscillating phase factors~\cite{kane1994randomness}.
Within the coupled-wire description of a moir\'{e} system, the momentum separation between neighboring wires is fixed by the lattice scale, $\Delta k=2\pi/\lambda$. This lattice determined momentum structure raises the possibility that oscillating phases in certain tunneling operators may be compensated by umklapp scattering. Whether such processes are actually allowed, however, depends on the microscopic relation between low-energy edge state operators and electronic degrees of freedom, which we analyze in the following sections.

{\it The decoupled $\nu=2/3$ state}---As a simple benchmark, we first examine the decoupled $nu=2/3$ state with diagonal $K$-matrix $K=\mathrm{diag}(3,3)$.
%
%
%
For this filling fraction, the Fermi momentum entering the coupled-wire description is fixed to $k_F^0=2\pi/3\lambda=b/3$~\cite{giamarchi2003quantum},
where $k_F^0$ is fixed by the electron density per moir\'e unit cell through a Luttinger-type relation, reflecting the quasi-one-dimensional representation of the underlying two-dimensional system~\cite{chou2019superconductor,hsu2023general}.

Within the coupled-wire construction, the quantum Hall phase is realized by introducing interwire interactions that gap out all bulk modes, leaving only gapless edge degrees of freedom.
For the present case, this can be achieved by the interaction operator~\cite{kane2002fractional}
\be
\mathcal{O}_{j}=\exp\Big(i(\varphi_j-\varphi_{j+2}+3(\theta_j+\theta_{j+2}))
\Big),
\ee
which enters the interacting Hamiltonian as
\be
H_{\rm int}=\sum_{j}\int dx (v\mathcal{O}_j(x)+{\rm H.c.}).
\ee

The resulting low energy theory contains two chiral edge modes $\phi_{1,2}$, which can be expressed in terms of the bosonic fields as~\cite{kane2002fractional}
\be
\phi_1=\frac{\varphi_1}{3}-\theta_1,
\quad \phi_2=\frac{\varphi_2}{3}-\theta_2,
\ee
which obey $[\phi_{\alpha},\phi_{\beta}]=-i\pi K^{-1}_{\alpha\beta}\sgn(x-x')$, with $K=\text{diag}(3,3)$~\cite{kane2002fractional}.
Charge conservation restricts inter-edge tunneling processes to operators of the form
$\exp(i3(\phi_1-\phi_2))$, which transfer three quasiparticles, carrying total charge $e$, from edge mode $\phi_2$ to edge mode $\phi_1$.
To analyze the momentum associated with this process, it is convenient to rewrite the tunneling operator in terms of electron operators~\cite{kane2002fractional}, which yields
%
\be
\exp(i3(\phi_1-\phi_2))=(\psi_{L,1}\psi^\dagger_{R,1}\psi_{L,1})(\psi_{L,2}\psi^\dagger_{R,2}\psi_{L,2})^{\dagger}.
\ee
From this expression, it is clear that the operator describes the tunneling of an electron from mode $\phi_2$ to mode $\phi_1$, and therefore carries a momentum difference $b=2\pi /\lambda$ between the two modes.
This momentum mismatch by a reciprocal lattice vector in a moir\'{e} system enables the corresponding tunneling due to lattice momentum conservation.

This demonstrates that the lattice-enabled inter-edge tunneling channels is a generic feature of the coupled-wire description in a moir\'{e} fractional Chern insulator.
In the present decoupled case, however, the corresponding operator $\exp(i3(\phi_1-\phi_2))$ is irrelevant in the renormalization group (RG) sense, with scaling dimension $\Delta=3$.
As we show below, this conclusion needs not hold in more intricate edge structures.
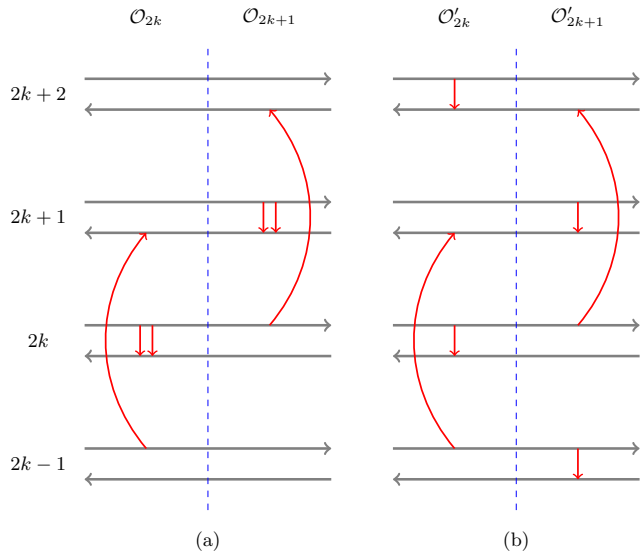
\begin{figure}[htbp]
  \centering
  \resizebox{1.0\linewidth}{!}{
  \input{fig2}}
  \caption{Schematic illustration of interwire tunneling processes leading to hierarchy $\nu=2/3$ states in the coupled-wire construction.
  In both panels, horizontal lines represent quantum wires and arrows indicate electron tunneling processes,  following the standard coupled-wire notation~\cite{kane2002fractional,teo2011luttinger}.
  }
  \label{fig:2}
\end{figure}

{\it Hierarchical $\nu=2/3$ state}---We now turn to the hierarchical state described by the $K$-matrix $K=\text{diag}(1,-3)$.
Within the coupled-wire construction, this topological phase admits multiple distinct microscopic realizations, which, as we show below, can lead to qualitatively different momentum structures for edge tunneling processes.

A standard coupled-wire realization of the hierarchical $\nu=2/3$ state is obtained by introducing the following set of interwire scattering operators~\cite{teo2011luttinger}:
\be \label{eq1:inter_wire_operator_1}
\mathcal{O}_{2k}&=&\exp\Big(i[\varphi_{2k-1}-\varphi_{2k+1}+4\theta_{2k}
\nn
&&+\theta_{2k+1}+\theta_{2k-1}]\Big),
\nn
\mathcal{O}_{2k+1}&=&\exp\Big(i[\varphi_{2k}-\varphi_{2k+2}+4\theta_{2k+1}
\nn
&&+\theta_{2k+2}+\theta_{2k}]\Big).
\ee
As shown in Refs.~\cite{teo2011luttinger,kane2002fractional}, these interactions gap out all bulk modes, leaving two gapless edge modes described by the bosonic fields $\Phi_{1,2}$
\be
\Phi_1&=&\varphi_1-\theta_1,
\nn
\Phi_2&=&\frac{1}{3}(-2\varphi_1+\varphi_2-2\theta_1-\theta_2),
\ee
which satisfy the commutation relations $[\Phi_{\alpha},\Phi_{\beta}]=-i\pi K^{-1}_{\alpha\beta}\sgn(x-x')$, with $K=\text{diag}(1,-3)$.
The most relevant charge conserving tunneling operator between the two edge modes is $\exp[i(\Phi_1+3\Phi_2)]$. Rewriting this operator in terms of bare electron operators yields~\cite{teo2011luttinger}
\be
\exp(i(\Phi_1+3\Phi_2))&=& \psi_{1R}^{\dagger}\psi_{1R}^{\dagger}\psi_{1L}\psi_{2L},
\ee
from which it follows that the process carries a momentum mismatch $4k^0_F-b=2\pi/(3\lambda)$,
%
rendering the corresponding tunneling process oscillatory and ineffective at long distances~\cite{kane1994randomness}~\footnote{While higher-order tunneling processes, i.e. $\exp(i3(\Phi_1+3\Phi_2))$, can in principle restore momentum conservation, they correspond to operators of higher scaling dimension and therefore strongly irrelevant.}.

This conclusion is not universal and depends on how the electronic degrees of freedom are embedded into the edge theory within the coupled-wire construction.
In the conventional approach, wires are grouped into pairs $(2k,2k+1)$, with each pair coupled to its neighbors by two interaction operators $\mathcal{O}_{2k}$ and $\mathcal{O}_{2k+1}$ of identical structure~\cite{teo2011luttinger}.
From the perspective of symmetry and locality, there is no fundamental requirement that these two operators take the same form.
Exploiting this freedom, we consider an alternative set of interwire interactions, as illustrated in Fig.~\ref{fig:2}(b),
\be \label{eq1:inter_wire_operator_2}
\mathcal{O}'_{2k}&=&\exp\Big(i[(\varphi_{2k-1}-\varphi_{2k+1})+2\theta_{2k}
\nn
&&+2\theta_{2k+2}+(\theta_{2k+1}+\theta_{2k-1})]\Big),
\nn
\mathcal{O}'_{2k+1}&=&\exp\Big(i[
(\varphi_{2k}-\varphi_{2k+2})+2\theta_{2k-1}
\nn
&&+2\theta_{2k+1}+(\theta_{2k+2}+\theta_{2k})]\Big).
\ee
With these interactions, one can again construct linear combinations of bosonic fields that gap the bulk while leaving two gapless edge modes~\cite{supplemental_material}, 
%
\be
\Phi'_1&=&\varphi_2-2\theta_1-\theta_2,
\nn
\Phi'_2&=&\frac{1}{3}(\varphi_1-2\varphi_2+3\theta_1),
\ee
which satisfy the same commutation relations as before with $K=\text{diag}(1,-3)$. 
Remarkably, in this realization the inter-edge tunneling operator $\exp(i(\Phi'_1+3\Phi'_2))$ can be written as~\cite{supplemental_material}
\be
\exp(i\Phi_1'+3i\Phi_2')=\psi^{\dagger}_{2R}\psi_{1R},
\ee
corresponding to the direct transfer of a bare electron between neighboring wires.
This simple microscopic structure stands in sharp contrast to the multi-electron tunneling processes encountered in the conventional realization.
The associated momentum mismatch is therefore $b=2\pi/\lambda$, which can be exactly compensated by a moir\'{e} reciprocal lattice vector via an umklapp process.

The origin of this difference can be made explicit by examining the microscopic form of the electron operators associated with the edge modes in the two realizations.
In the first realization, the edge electron operators take the form $\exp(i\Phi_1) \sim \psi_{1L}$ and $\exp(-i3\Phi_2) \sim  \psi_{1R}\psi_{1R}\psi^\dagger_{2L}$, whereas in the second realization they are given by $\exp(i\Phi'_1) \sim  \psi_{1L}\psi_{2L}\psi^\dagger_{1R}$ and $\exp(-i3\Phi'_2) \sim \psi_{2R}\psi_{1L}\psi_{2L}\psi^\dagger_{1R}\psi^\dagger_{1R}$.
Although both constructions yield the same $K$-matrix structure, the microscopic embedding of electronic degrees of freedom into the edge theory is qualitatively different in the two cases.
Interestingly, previous works conjectured that distinct choices of interwire interactions at $\nu=2/3$ may correspond to different microscopic fractional quantum Hall states, such as the particle-hole conjugate of the Laughlin $\nu=1/3$ state and the Halperin $(112)$ state~\cite{fuji2019quantum} .

Since the interaction terms [Eqs.~(\ref{eq1:inter_wire_operator_1},\ref{eq1:inter_wire_operator_2})] in the two realizations have identical scaling dimensions, the RG analysis alone does not favor one realization over the other.
To further contrast the two microscopic realizations, we examine their momentum structure in a finite system consisting of $2\ell+2$ coupled-wires.
In the standard realization, electron operators on opposite edges differ by momenta that include fractional offsets, such as $b(2\ell+2+2/3)$.
%
In the alternative realization the corresponding momentum differences are fixed by the geometric separation between edge modes and given by integer multiples of $b$ (see the supplementary material~\cite{supplemental_material}).

This difference becomes especially clear when the effective momentum scale $b$ is formally interpreted as arising from a background magnetic field, as in the original coupled-wire construction of the fractional quantum Hall effect~\cite{kane2002fractional,teo2011luttinger}.
In this sense, the alternative realization yields a more transparent momentum bookkeeping, with edge-to-edge momentum differences that directly track their physical separation~\cite{supplemental_material}.
In a moir\'{e} fractional Chern insulator, the effective momentum scale $b$ does not correspond to a real electromagnetic field. 
Hence the above considerations can not determine which microscopic realization is selected. Notwithstanding they imply qualitatively different edge dynamics, as we discuss below.

{\it Umklapp-driven KFP fixed point}---
These distinct momentum structures have direct consequences for the edge dynamics.
In the second coupled-wire realization, the momentum mismatch of the inter-edge tunneling operator equals a reciprocal lattice vector $2\pi/\lambda$. As a result, the edge theory admits an additional umklapp process forbidden in the continuum.
One such leading process is described by the action $S_{u}=(u/4)\int dxd\tau \exp[i(\Phi'_1+3\Phi'_2)]+\text{H.c.}$, 
where the momentum mismatch is exactly compensated by a reciprocal lattice vector.

Now, it is convenient to introduce the charge and neutral fields $\phi_{\rho}=\sqrt{3/2}(\Phi'_1+\Phi'_2)$, $\phi_{\sigma}=\sqrt{1/2}(\Phi'_1+3\Phi'_2)$, in terms of which the edge action takes the form~\cite{kane1994randomness}
\be 
S_{\rho}&=&\frac{1}{4\pi}\int dx d\tau \partial_x\phi_{\sigma} (i\partial_\tau+v_{\rho}\partial_x)\phi_{\rho},
\nn 
S_{\sigma}&=&\frac{1}{4\pi}\int dx d\tau \partial_x\phi_{\sigma}(-i\partial_\tau+v_{\sigma}\partial_x)\phi_{\sigma}
\nn
&&+\frac{u}{2}\int dxd\tau \cos(\sqrt{2}\phi_\sigma),
\nn 
S_{\rm pert}&=&v\frac{1}{4\pi}\int dx d\tau \partial_x\phi_{\rho}\partial_x \phi_{\sigma}.
\ee

Following an analysis closely paralleling that of KFP 
~\cite{supplemental_material,kane1994randomness}, the tunneling term in $S_{\sigma}$ introduces an intrinsic momentum scale of order $u$, which becomes manifest when the neutral sector is described in a basis adapted to the regime in which the tunneling term dominates.
Since the operator $\exp(i(\Phi'_1+3\Phi'_2))$ is relevant for initial conditions within the basin of attraction of the KFP fixed point~\cite{kane1994randomness}, the coupling $u$ grows under RG flow.
In this regime, forward-scattering processes $S_{\rm pert}$, which involve only small momentum transfer, become kinematically ineffective at low energies~\cite{supplemental_material}.
We therefore conclude that, although the operator $\partial_x\phi_{\rho}\partial_x\phi_{\sigma}$ is marginal by power counting, its contribution is dynamically suppressed at low energies by the lattice-enabled umklapp process, and the resulting infrared theory is governed by the well-established KFP fixed point.
Consequently, the two microscopic realizations discussed above lead to qualitatively different edge dynamics: only those in which the inter-edge tunneling operator is rendered momentum-conserving by lattice umklapp processes flow to the KFP fixed point, while the others remain outside this equilibrated regime.
This distinction can be probed experimentally, for example through two-terminal conductance or quantum point contact measurements~\cite{kane1994randomness,kane1995impurity,manna2025multiple}, which distinguish between equilibrated and non-equilibrated edge regimes.

{\it Conclusion}---In summary, we have shown that the momentum structure of fractional edge modes in moir\'{e} fractional Chern insulators depends sensitively on their microscopic realization, even when the $K$-matrix is fixed.
Distinct microscopic realizations lead to qualitatively different edge dynamics: while some prohibit inter-edge tunneling due to momentum mismatch, others admit lattice-enabled umklapp processes that stabilize the KFP fixed point even in the absence of disorder.
Our results thus identify a lattice-enabled mechanism by which edge equilibration becomes realization-dependent in moir\'{e} systems.
These results highlight the role of lattice momentum constraints in reshaping fractional edge physics and suggest that moir\'{e} platforms provide a fertile setting for exploring lattice-driven edge phenomena beyond the continuum quantum Hall paradigm.

An interesting open question concerns whether the lattice-stabilized KFP fixed point identified here can be experimentally distinguished from the disorder-driven scenario originally discussed by KFP~\cite{kane1994randomness,kane1995impurity}.
Although the zero-temperature fixed point is the same, the physical mechanism suppressing the charge-neutral coupling differs qualitatively: disorder averaging in the original KFP scenario versus kinematic momentum mismatch in the present lattice setting.
It would be interesting to explore whether this distinction manifests itself at finite temperature, for example in the dynamics or equilibration of neutral modes.

From a microscopic perspective, it would be interesting to understand whether specific realizations of moir\'{e} electronic networks---such as domain-wall or quantum-network states---naturally favor one microscopic edge realization over another,and therefore select between distinct edge dynamical regimes.
Addressing this question would require a more detailed microscopic or numerical analysis, which we leave for future work.
Finally, the framework developed here suggests a broader applicability to other fractional edge theories in lattice settings, e.g. proposed in recent works~\cite{tam2026quantized}.

{\it Acknowledgments}---We thank T. Senthil and Taige Wang for inspirations and helpful discussions. XYS is supported by the Croucher innovation award. H.C.P. acknowledges support from the Hong Kong Research Grants Council through project C7037-22GF.

\bibliography{E3}

\onecolumngrid
\onecolumngrid

\newpage
\clearpage

\vspace{1cm}
\begin{center}
\textbf{\large Supplemental material: Moir\'{e} driven edge reconstruction in Fractional quantum anomalous Hall states}
\end{center}

\setcounter{secnumdepth}{3}
\setlength\parindent{0pt}

\setcounter{equation}{0}
\setcounter{figure}{0}
\setcounter{table}{0}
\setcounter{page}{1}

\makeatletter
\renewcommand{\theequation}{S\arabic{equation}}
\renewcommand{\thefigure}{S\arabic{figure}}
\renewcommand{\thetable}{S\arabic{table}}
\renewcommand{\thesection}{S\arabic{section}}
\newcommand{\lpartial}{\overleftarrow{\partial}}
\newcommand{\rpartial}{\overrightarrow{\partial}}
\newcommand{\lnabla}{\overleftarrow{\nabla}}
\newcommand{\rnabla}{\overrightarrow{\nabla}}

\section{Microscopic realizations of the hierarchical \texorpdfstring{$\nu=2/3$}{nu=2/3} edge in the coupled-wire construction}

We first briefly review the standard coupled-wire construction of the hierarchical
$\nu=2/3$ state following Ref.~\cite{teo2011luttinger}.
The basic interwire interaction takes the form
\be
\mathcal{O}_j=\exp\Big(i[(\varphi_{j-1}-\varphi_{j+1})+4\theta_j+(\theta_{j+1}+\theta_{j-1})]
\Big)
\ee
The wires are grouped into pairs $(2k,2k+1)$, with each pair coupled to its neighbors by two tunneling operators
\be
\mathcal{O}_{2k}&=&\exp\Big(i[(\varphi_{2k-1}-\varphi_{2k+1})+4\theta_{2k}+(\theta_{2k+1}+\theta_{2k-1})]\Big)
\nn
\mathcal{O}_{2k+1}&=&\exp\Big(i[(\varphi_{2k}-\varphi_{2k+2})+4\theta_{2k+1}+(\theta_{2k+2}+\theta_{2k})]\Big)
\ee
Introducing the bosonic fields $\tilde{\phi}^{R/L}_{k,a}$ as in Ref.~\cite{teo2011luttinger}, the tunneling operators gap all bulk modes, leaving two chiral edge modes.
After an $SL(2,\mathbb{Z})$ transformation, the edge theory is described by bosonic fields
\be
\Phi_1&=&\varphi_1-\theta_1
\nn
\Phi_2&=&\frac{1}{3}(-2\varphi_1+\varphi_2-2\theta_1-\theta_2)
\ee
which obey $[\Phi_\alpha(x),\Phi_\beta(x')]=-i\pi K^{-1}_{\alpha\beta}\sgn(x-x')$ with $K=\mathrm{diag}(1,-3)$.

The most relevant charge-conserving tunneling operator between the two edge modes is
$\exp(i\Phi_1+3i\Phi_2)$.
Expressed in terms of microscopic electron operators, one finds
\be
\exp(i\Phi_1+3i\Phi_2)
=
\psi_{1R}^{\dagger}\psi_{1R}^{\dagger}\psi_{1L}\psi_{2L},
\ee
corresponding to a multi-electron tunneling process that carries a fractional momentum mismatch
$4k_F^0-b=2\pi/(3\lambda)$.
As discussed in the main text, this mismatch cannot be compensated by a single moir\'e reciprocal lattice vector and therefore renders the corresponding tunneling operator ineffective at low energies.

\hfill

We now consider an alternative set of interwire interactions,
\be
\mathcal{O}'_{2k}
&=&\exp\Big(i[(\varphi_{2k-1}-\varphi_{2k+1})+2\theta_{2k}+2\theta_{2k+2}+(\theta_{2k+1}+\theta_{2k-1})]
\Big)
\nn
\mathcal{O}'_{2k+1}&=&\exp\Big(i[
(\varphi_{2k}-\varphi_{2k+2})+2\theta_{2k-1}+2\theta_{2k+1}+(\theta_{2k+2}+\theta_{2k})
]
\Big)
\ee
which are equally local and symmetry-allowed.
Defining a different set of bosonic variables $\tilde{\phi}^{R/L}_{k,a}$,
\be
\tilde{\phi}^R_{k,1}&=&\varphi_{2k-1}+\theta_{2k-1}+2\theta_{2k}
\nn
\tilde{\phi}_{k,1}^L&=&\varphi_{2k-1}-\theta_{2k-1}-2\theta_{2k}
\nn
\tilde{\phi}^R_{k,2}&=&\varphi_{2k}+\theta_{2k}+2\theta_{2k-1}
\nn
\tilde{\phi}^L_{k,2}&=&\varphi_{2k}-\theta_{2k}-2\theta_{2k-1}
\ee
we can rewirte the tunneling terms $\mathcal{O}'_{2k}$ and $\mathcal{O}'_{2k+1}$ as
\be
\mathcal{O}_{2k}&=&\exp\Big(i(\tilde{\phi}^R_{k,1}-\tilde{\phi}^L_{k+1,1})
\Big)
\nn
\mathcal{O}_{2k+1}&=&\exp\Big(i(\tilde{\phi}^{R}_{k,2}-\tilde{\phi}_{k+1,2}^L)
\Big)
\ee
and one again finds that the bulk is fully gapped and the resulting topological order is described by the same $K=\mathrm{diag}(1,-3)$ matrix.
At the edge, the remaining chiral modes may be written as
\be
\Phi'_1&=&\varphi_2-2\theta_1-\theta_2
\nn
\Phi'_2&=&\frac{1}{3}(\varphi_1-2\varphi_2+3\theta_1)
\ee
which satisfy the same $K$-matrix algebra as in the standard construction.
However, in this realization the simplest inter-edge tunneling operator takes the form
\be
\exp(i\Phi_1'+3i\Phi_2')=\exp(\varphi_1-\varphi_2+\theta_1-\theta_2)=\psi^{\dagger}_{2R}\psi_{1R}
\ee
corresponding to the direct tunneling of a single electron between two neighboring wires at the edge.
The associated momentum mismatch is $b=2\pi/\lambda$, which coincides with a reciprocal lattice vector of the moir\'e superlattice and can therefore be absorbed by an umklapp process.

\section{Edge-to-edge quasiparticle tunneling in a finite coupled-wire system and quantum anomaly}

We now examine the topological transfer of quasiparticles between the two edges in a finite coupled-wire system consisting of $2\ell+2$ wires.

For both microscopic realizations discussed in the main text, one can explicitly construct the bosonic fields describing the edge modes on the top boundary, denoted schematically by $\tilde{\Phi}_{1,2}$.
As in the bottom edge discussed previously, the operators $e^{i\tilde{\Phi}_1}$ and $e^{i\tilde{\Phi}_2}$ create quasiparticles of charge $e$ and $-e/3$, respectively.
While the individual operator $e^{i\tilde{\Phi}_2}$ ($e^{i\Phi_2}$) is not by itself a physical electron operator, quasiparticle tunneling between the two edges can be realized by local string operators built from interwire scattering terms, as shown in Ref.~\cite{teo2011luttinger}.

\begin{figure}[htbp]
  \centering
  \resizebox{0.2\linewidth}{!}{
  \input{fig3}}
  \caption{Schematic droplet picture of a finite-width $\nu=2/3$ FQHE~\cite{wang2013edge,wen2004quantum}.
  The red arrows illustrate a topological pumping process that transfers a net charge $2e/3$ quasiparticle from the bottom edge to the top edge, composed of one $e$ mode and one $-e/3$ mode.
  }
  \label{fig:3}
\end{figure}
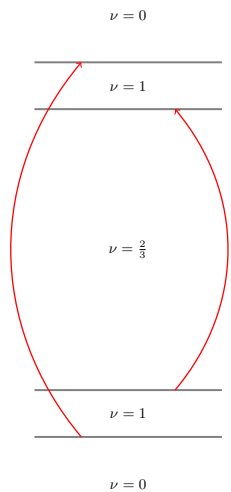

The transfer of a net charge-$2/3$ quasiparticle from the bottom to the top edge may be represented by a composite operator of the form
\be
\mathcal{O}_{\mathrm{tb}}&=&\exp\big[i(\Phi_1+\Phi_2)-i(\tilde{\Phi}_1+\tilde{\Phi}_2)\big]
\nn
\mathcal{O}'_{\mathrm{tb}}&=&\exp\big[i(\Phi'_1+\Phi'_2)-i(\tilde{\Phi}'_1+\tilde{\Phi}'_2)\big],
\ee
where $\Phi_{1,2}$($\Phi'_{1,2}$) and $\tilde{\Phi}_{1,2}$($\tilde{\Phi}'_{1,2}$) denote the edge fields on the bottom and top boundaries, respectively.
Importantly, although the microscopic structure of this operator depends on the chosen realization, the total momentum transferred in this process is the same in both cases and is given by
\be
\label{eq:momentumchange}
\Delta p = \frac{2}{3}(2\ell) b + 2b
= \frac{2}{3}(2\ell)\frac{2\pi}{\lambda} + 2\frac{2\pi}{\lambda}.
\ee

This result is not accidental. It reflects the fact that the tunneling process corresponds to a topological pump of a net charge-$2/3$ quasiparticle, composed of one $e$ mode and one $-e/3$ mode.
This result is consistent with the Oshikawa argument~\cite{oshikawa2000commensurability,oshikawa2000topological}, which relates momentum pumping to the filling profile across the system. Consider the quasi-1D limit where the horizontal direction obeys the periodic boundary condition, i.e. forming a ring and thread a $2\pi$ flux along the periodic direction, this will pump a $2e/3$ charged object from the bottom to the top of the ring. This flux threading can be compensated by a large gauge transform on the ring induced by the operator $\exp[i\frac{2\pi}{L_x}\int \hat \rho(x,y)\hat x dxdy]$, which take the Hamiltonian back to the original form. However this operator will change the total momentum $\hat P_x\rightarrow \hat P_x+\frac{2\pi}{L_x}\langle\int \rho(x,y)dx dy\rangle$. Under the circumstance of uniform density in the bulk and between the two edge modes as depicted in Fig.~\ref{fig:3}, the momentum change of the process amounts to the expression Eq.~\eqref{eq:momentumchange}. Hence the coupled-wire construction agrees with the anomaly structure of the fractional quantum anomalous Hall state.

By comparing this momentum transfer $\Delta p$ with the droplet picture illustrated in Fig.~\ref{fig:3}, one finds that the two e modes are separated by $2\ell+2$ wires, while the two $-e/3$ modes are separated by $2\ell$ wires.
If one formally interprets the momentum scale $b$ as arising from a background magnetic field, as in the original coupled-wire construction of the FQHE~\cite{kane2002fractional,teo2011luttinger}, this implies momentum differences of $(2\ell+2)b$ and $2\ell b$ for the corresponding electron operators. Remarkably, this structure is realized explicitly in the second microscopic construction discussed in the main text, where the edge momentum mismatches directly reflect these separations.
By contrast, this correspondence is obscured in the first realization, where fractional momentum offsets appear. 

However, in a moir\'{e} fractional Chern insulator the effective scale $b$ does not correspond to a real electromagnetic field. Consequently, while this interpretation provides useful physical intuition, the present analysis does not uniquely determine which microscopic realization is ultimately selected.

\section{Umklapp-induced suppression of charge-neutral coupling}

The Euclidean action for a hierarchical $\nu=2/3$ edge state is given by
\be \label{eqs:Euclidean_action_1}
S_0=\frac{1}{4\pi}\int dxd\tau \Big((\partial_x\phi_i)K_{ij}(i\partial_{\tau}\phi_j)+v_{ij}(\partial_x\phi_i)(\partial_x \phi_j)
\Big)
\ee
where $K=\mathrm{diag}(1,-3)$ and $v_{ij}$ is a symmetric, positive-definite velocity matrix.
We now consider electron tunneling between the two edge modes $\phi_1$ and $\phi_2$.
Charge conservation restricts inter-edge tunneling to operators of the form
$\exp[i(m\phi_1+n\phi_2)]$ with $m-n/3=0$.
The most relevant such operator is $e^{i(\phi_1+3\phi_2)}$, which transfers three quasiparticles, carrying total charge $e$, from mode $2$ to mode $1$~\cite{kane1994randomness}.
The corresponding contribution to the action is
\be
S_1
=\frac{u}{4}\int d x d\tau 
\Big[e^{i(\phi_1+3\phi_2)}+\mathrm{H.c.}\Big]
=\frac{u}{2}\int d x d\tau 
\cos(\phi_1+3\phi_2).
\ee
Importantly, this operator appears without an oscillatory spatial phase factor.
As discussed in the main text, in the second coupled-wire realization the momentum mismatch associated with this process equals a reciprocal lattice vector $2\pi/\lambda$ and can therefore be absorbed by a moir\'e umklapp process.
Equivalently, in accordance with the momentum bookkeeping interpretation discussed in the main text, the apparent momentum shift reflects the lattice momentum structure intrinsic to the coupled-wire description and does not lead to an oscillatory phase in the low-energy effective theory.

\hfill

To analyze the effect of the tunneling term, it is convenient to introduce charge and neutral modes via
\be
\phi_{\rho}=\sqrt{\frac{3}{2}}(\phi_1+\phi_2),
\qquad
\phi_{\sigma}=\sqrt{\frac{1}{2}}(\phi_1+3\phi_2),
\ee
where $\phi_\rho$ carries the electric charge and $\phi_\sigma$ is electrically neutral.
In terms of these fields, the total action $S=S_0+S_1$ may be written as
\be
S=S_\rho+S_\sigma+S_{\rm pert},
\ee
with 
\be
S_{\rho}&=&\frac{1}{4\pi}\int dx d\tau \partial_x\phi_{\sigma} (i\partial_\tau+v_{\rho}\partial_x)\phi_{\rho} 
\nn 
S_{\sigma}&=&\frac{1}{4\pi}\int dx d\tau \partial_x\phi_{\sigma}(-i\partial_\tau+v_{\sigma}\partial_x)\phi_{\sigma}+\frac{u}{2}\int dxd\tau \cos(\sqrt{2}\phi_\sigma)
\nn 
S_{\rm pert}&=&v\frac{1}{4\pi}\int dx d\tau \partial_x\phi_{\rho}\partial_x \phi_{\sigma} 
\ee
The velocities $v_\rho$, $v_\sigma$, and $v$ are nonuniversal functions of the parameters in $S_0$ and will be treated as independent constants.

The central question is the fate of the charge--neutral coupling $S_{\rm pert}$ once the tunneling term in $S_\sigma$ becomes important at low energies.
The leading renormalization-group (RG) flow equation for $u$ is
\be
\frac{d u}{d \ell}=(2-\Delta)u,
\ee
where $\Delta$ is the scaling dimension of the operator $e^{i(\phi_1+3\phi_2)}$.
A controlled description of this regime was developed by Kane, Fisher, and Polchinski~\cite{kane1994randomness,kane1995impurity}.
They showed that the quadratic part of the neutral-sector action $S_\sigma$ respects SU(2) symmetry and can be mapped exactly onto a theory of chiral fermions with identical low-energy physics.
Rather than repeating this standard construction, we directly adopt the resulting fermionic representation.
Specifically, the neutral sector may be written in terms of a two-component chiral fermion $\psi=(\psi_1,\psi_2)^T$ as
\be
S_{\psi} =\int d x d\tau \psi^\dagger(x,\tau)
\Big(\partial_\tau+i v_\sigma\partial_x+u\sigma_x
\Big)\psi(x,\tau)
\ee
where $\sigma_\mu$ are Pauli matrices acting in the internal space of $\psi$.
The $u$ term can be eliminated from $S_\sigma$ by a unitary $SU(2)$ gauge transformation,
\be
\tilde{\psi}(x)=U(x)\psi(x),
\qquad
U(x)=\exp \Big[-i\frac{u}{v_\sigma}\sigma_x x\Big].
\ee
In terms of $\tilde{\psi}$, the neutral-sector action $S_{\sigma}$ becomes that of free chiral fermions,
\be
S_{\tilde{\psi}} = \int d x d\tau
\tilde{\psi}^\dagger(x,\tau)
(\partial_\tau+i v_\sigma\partial_x)
\tilde{\psi}(x,\tau).
\ee

\hfill

We now examine the perturbation $S_{\rm pert}$, which couples the charge and neutral sectors.
In fermionic variables, it takes the form
\be
S_{\rm pert}= v\frac{1}{4\pi}\int d x d\tau
\partial_x\phi_\rho \psi^\dagger\sigma_z\psi
\ee
After the unitary transformation, this becomes
\be
S_{\rm pert}
= v\frac{1}{4\pi}\int d x d\tau \partial_x\phi_\rho \tilde{\psi}^\dagger \big[U(x)\sigma_z U^\dagger(x)\big]
\tilde{\psi}
\ee
with
\be
U(x)\sigma_z U^\dagger(x)
=
\cos\Big(\frac{2u}{v_\sigma}x\Big)\sigma_z
-\sin\Big(\frac{2u}{v_\sigma}x\Big)\sigma_y.
\ee
The operator $\partial_x\phi_\rho \tilde{\psi}^{\dagger}\tilde{\psi}$ has scaling dimension two and would be marginal if its coefficient were spatially uniform.
However, for initial conditions such that the vertex operator $\exp(i\phi_1+3i\phi_2)$ is relevant, the RG flow drives the coupling $u$ to grow under coarse graining, eventually reaching a regime in which $u$ becomes parametrically large.

In this regime, the charge-neutral coupling $S_{\rm pert}$ acquires a rapidly oscillating spatial dependence with wavevector $2u/v_\sigma$.
Since the low-energy effective theory contains only modes with momenta much smaller than $2u/v_\sigma$, this perturbation cannot connect low-energy states desecibed by $S_{\rho}+S_{\tilde{\psi}}$ and is therefore kinematically suppressed in the infrared.
As a result, the perturbation term $S_{\rm pert}$ becomes ineffective at low energies and does not modify the infrared structure of the effective theory.

This mechanism should be contrasted with the disorder-induced irrelevance discussed in Ref.~\cite{kane1994randomness}, where the suppression of $S_{\rm pert}$ arises from spatial randomness rather than from momentum mismatch.
Consequently, the charge-neutral coupling becomes inactive at low energies, and the resulting infrared description is consistent with the Kane-Fisher-Polchinski fixed point characterized by decoupled charge and neutral modes.

\section{Quantum point contact conductance exponents}

As pointed out by Kane, Fisher, and Polchinski~\cite{kane1994randomness,kane1995impurity}, tunneling through a quantum point contact (QPC) provides a sensitive probe of the structure of $\nu=2/3$ edge states. 
At low temperatures, the conductance scales as
\be
G(T)\sim T^{\alpha},
\ee
where the exponent $\alpha$ is determined by the scaling dimension of the most relevant charge $Q=1$ tunneling operator~\cite{kane1994randomness,kane1995impurity}.
Within Euclidean action for a hierarchical $\nu=2/3$ edge state [Eq.~(\ref{eqs:Euclidean_action_1})], the scaling dimension of a vertex operator $\exp(i( m_1\phi_1+m_2\phi_2))$ is given by~\cite{tam2026quantized}
\be
\Delta(m_1,m_2)
=\Big(m_1^2+\frac{m_2^2}{3}\Big)\cosh 2\theta
-\frac{2}{\sqrt{3}}m_1m_2\sinh 2\theta,
\ee
where the parameter $\theta$ is determined by the interaction matrix $v_{ij}$ via
\be
\tanh 2\theta =\frac{6v_{12}}{\sqrt{3}(3v_{11}+v_{22})}.
\ee
The allowed tunneling operators satisfy the charge constraint
\be
Q = m_1 - \frac{m_2}{3} = 1.
\ee
In a generic (non-KFP) edge theory, the scaling dimension depends explicitly on the interaction parameters through $\theta$, and different allowed operators may compete to be the most relevant. 
The most relevant operator therefore has a scaling dimension $\Delta_{\min}$ that varies with $v_{ij}$, leading to a conductance exponent
\be
\alpha = 2(2\Delta_{\min}-1),
\ee
which is non-universal.

\hfill

By contrast, at the Kane-Fisher-Polchinski fixed point, the charge and neutral sectors decouple at low energies, and the scaling dimension of the charge $Q=1$ tunneling operator becomes universal,
\be
\Delta_{\min}=1,
\quad
\alpha=2,
\ee
leading to the well-known result~\cite{kane1994randomness}
\be
G(T)\sim T^2.
\ee

This distinction implies that the presence or absence of the KFP fixed point can be directly probed through QPC measurements, which distinguish between interaction-dependent and universal conductance scaling.

\end{document}

%% file: fig2.tex
\begin{tikzpicture}
\draw[<-,gray,very thick](-4,-4) -- (0,-4);
\draw[->,gray,very thick](-4,-3.5) -- (0,-3.5);
\draw[<-,gray,very thick](-4,-2) -- (0,-2);
\draw[->,,gray,very thick](-4,-1.5) -- (0,-1.5);
\draw[<-,gray,very thick](-4,0) -- (0,0);
\draw[->,,gray,very thick](-4,0.5) -- (0,0.5);
\draw[<-,gray,very thick](-4,2) -- (0,2);
\draw[->,gray,very thick](-4,2.5) -- (0,2.5);
\draw[->, red,thick] (-3,-3.5) to[bend left=40] (-3,0);
\draw[->, red,thick] (-1,-1.5) to[bend right=40] (-1,2);
\draw[->,red,thick] (-3.1,-1.5)--(-3.1,-2);
\draw[->,red,thick] (-2.9,-1.5) -- (-2.9,-2);
\draw[->,red,thick] (-1.1,0.5)--(-1.1,0);
\draw[->,red,thick] (-0.9,0.5) -- (-0.9,0);
\node at (-4.75,-3.75) {$2k-1$};
\node at (-4.75,-1.75) {$2k$};
\node at (-4.75,0.25) {$2k+1$};
\node at (-4.75,2.25) {$2k+2$};
\node at (-3,3.5) {$\mathcal{O}_{2k}$};
\node at (-1,3.5) {$\mathcal{O}_{2k+1}$};
\draw[dashed, blue] (-2,-4.5) -- (-2,3);

\draw[<-,gray,very thick](1,-4) -- (5,-4);
\draw[->,gray,very thick](-4+5,-3.5) -- (0+5,-3.5);
\draw[<-,gray,very thick](-4+5,-2) -- (0+5,-2);
\draw[->,,gray,very thick](-4+5,-1.5) -- (0+5,-1.5);
\draw[<-,gray,very thick](-4+5,0) -- (0+5,0);
\draw[->,,gray,very thick](-4+5,0.5) -- (0+5,0.5);
\draw[<-,gray,very thick](-4+5,2) -- (0+5,2);
\draw[->,gray,very thick](-4+5,2.5) -- (0+5,2.5);
\draw[->, red,thick] (-3+5,-3.5) to[bend left=40] (-3+5,0);
\draw[->, red,thick] (-1+5,-1.5) to[bend right=40] (-1+5,2);
\draw[->,red,thick] (-3+5,2.5)--(-3+5,2);
\draw[->,red,thick] (-3+5,-1.5) -- (-3+5,-2);
\draw[->,red,thick] (-1+5,-3.5)--(-1+5,-4);
\draw[->,red,thick] (-1+5,0.5) -- (-1+5,0);
\node at (-3+5,3.5) {$\mathcal{O}'_{2k}$};
\node at (-1+5,3.5) {$\mathcal{O}'_{2k+1}$};
\draw[dashed, blue] (-2+5,-4.5) -- (-2+5,3);

\node at (-2,-5) {(a)};
\node at (3,-5) {(b)};

\end{tikzpicture}

%% file: fig3.tex
\begin{tikzpicture}
\draw[gray,very thick](-4,-4) -- (0,-4);
\draw[gray,very thick](-4,-3) -- (0,-3);
\draw[gray,very thick](-4,3) -- (0,3);
\draw[gray,very thick](-4,4) -- (0,4);
\draw[->, thick,red] (-3,-4) to[bend left=40] (-3,4);
\draw[->, thick,red] (-1,-3) to[bend right=40] (-1,3);
%
%
\node at (-2,3.5) {$\nu=1$};
\node at (-2,-5) {$\nu=0$};
\node at (-2,0) {$\nu=\frac{2}{3}$};
\node at (-2,-3.5) {$\nu=1$};
\node at (-2,5) {$\nu=0$};

\end{tikzpicture}